\documentclass[12pt,A4]{article}
\usepackage{array}
\usepackage{amsmath}
\newcommand{\Chi}{\mathrm X}
\newcommand{\be}{\begin{equation}}
\newcommand{\ee}{\end{equation}}
\newcommand{\bea}{\begin{eqnarray}}
\newcommand{\eea}{\end{eqnarray}}
\newcommand{\m}{\mathbf}
\def\la{\mathrel{\mathpalette\fun <}}

\def\fun#1#2{\lower3.6pt\vbox{\baselineskip0pt\lineskip.9pt
\ialign{$\mathsurround=0pt#1\hfil##\hfil$\crcr#2\crcr\sim\crcr}}}

\begin{document}
 \title{RIGHT-LEFT ASYMMETRY OF RADIATION FROM FISSION}
\author{F.F.Karpeshin \\
\em Fock Institute of Physics, St. Petersburg State University\\
RU-198504 St. Petersburg, Russia}

\maketitle

\begin{abstract}
The effect of the right-left asymmetry is considered in the predicted earlier
electric dipole radiation from fission fragments arising due to the
Strutinski---Denisov induced polarisation mechanism.
The magnitude of the asymmetry parameter is on the level of  10$^{-3}$.
That is in agreement with the recent experimental data on the radiative
ROT effect in $^{235}$U fission induced by cold polarised neutrons.
\end{abstract}



\newpage
\section{Introduction}

The ROT effect of the right-left asymmetry in the emission of light charged
particles in ternary fission
with respect to the plane formed by the directions of the spin of
the fissile nucleus and the fission axis has been found some time ago~\cite{ROT}.
Very recently such measurements were performed with $\gamma$ quanta from
usual $^{235}$U fission induced by cold polarised neutrons~\cite{dan}.
Strong right-left asymmetry has been detected in \cite{dan} at the level
of 10$^{-3}$.
The electric dipole radiation from fragments was suggested in \cite{dub93}
on the basis of the
Strutinski---Denisov induced polarisation mechanism~\cite{D6,D7}.
The probability of the emission has been obtained of the order of
\be
N_\gamma^{E1} = 0.008\,\mbox{fission}^{-1}\;.   \label{Ng}
\ee
This value is high enough to explain the results \cite{dan}.
The radiation is emitted from fragments
before the neutron emission due to snapping back the nuclear surface
within a time interval $\tau_{dis}$ which is determined by dissipation
of the collective energy. According to \cite{D5}, $\tau_{dis}\la 10^{-19}$~s.
Let us examine consequence of this mechanism for the ROT effect with
$\gamma$ quanta.

   \section{Physical premises}

 Experiment shows that the fragments are formed partly aligned after scission
in the plane perpendicular to the fission axis, with the average spin of
$I$ = 7 -- 8. The effect of alignment was observed in the angular
distribution of the emitted radiation \cite{skars} and conversion muons
from the prompt fission fragments~\cite{np97}.
The spins of the fragments can be parallel or antiparallel to each other.
In the last case, their total spin is compensated by the large angular
momentum of the relative motion of the fragments.
At the moment of scission, the nascent fragments have a pear-like form
with the noses directed towards the point of the rupture.
Wavefunction of a fragment can be presented as follows~\cite{BM}:
\be
\psi(\m r) = \left(\frac{2I+1}{16\pi^2}\right)^{1/2}
\begin{cases}
D^I_{MK}\phi_K(\m r') + (-1)^{I+K}D^I_{M\ -K}\phi_{-K}(\m r') &
\text{for $\pi$=1,}\\
i\left[D^I_{MK}\phi_K(\m r') -(-1)^{I+K}D^I_{M\ -K}\phi_{-K}(\m r')\right] &
\text{for $\pi$=-1.}
\end{cases}
\label{grg1}
\ee
$K$ is thus a good quantum number. Therefore,
evolution of the nuclear surface occurs under holding axial  symmetry
of the fragment. The symmetry axis rotates in the plane perpendicular
to the spin of the fissile nucleus~\cite{ka84}.
Let the angular distribution in the intrinsic system be
\be
\chi(\theta')= 1 - a \cos^2\theta'\;\;.     \label{grg2}
\ee
And let the angle of the symmetry axis at the moment of emission
against the quantization axis $z$ be
\be
\Omega(t) = \omega t\;.
\ee
Then the angular distribution in the laboratory frame reads
\be
\Chi(\theta, t) =  \chi\left(\theta - \Omega(t)\right)\;.
\ee
Integrating the last expression over time and taking into account
the time of relaxation due to dissipation of the collective energy
 $\tau_{dis}\equiv 1/\lambda$, we arrive at the angular distribution
in the laboratory system
\bea
\Chi(\theta) = \int_0^\infty \chi\left(\theta-\Omega(t)\right)
\gamma \exp(-\gamma t) \ dt = \\
= \int_0^\infty \left[1-a\cos^2(\theta-  \omega t)\right] \gamma
\exp(-\gamma t)\ dt= \\
= 1 - a\cos^2(\theta-\delta)\;,      \label{grg6}
\eea
where $\tan 2\delta = 2 \frac\omega\lambda$, that is in the laboratory
frame the distribution remains the same in the form,
but with the shifted by angle $\delta$ argument.
Reversal of the polarisation of the incoming neutrons is
equivalent to the reversal of the direction of the rotation of the
fissile nucleus and the fragments. This leads to a replacement
$\delta \to -\delta$. Therefore, we arrive
by means of eq. (\ref{grg6}) at the following expression for
the asymmetry parameter
\bea
R(\theta)=\frac{a[\cos^2(\theta+\delta)-\cos^2(\theta-\delta)]}
{2-a[\cos^2(\theta+\delta)+\cos^2(\theta-\delta)]} =\\=
-\frac12 a\sin 2\theta \sin 2\delta\;. \label{grg7}
\eea
Let us consider two opposite limiting cases of (\ref{grg6}).

	1) $\omega \ll \lambda$. The shifting angle remains small, as the
fragment has not enough time to accomplish the revolution before radiating:
\be
\delta \approx \frac\omega  \lambda\;.
\ee
Actually, this case is usually considered for explanation of the ROT
effect with light charged particles~\cite{ROT}, with rotation of the
fission axis instead of rotation of the symmetry axis of a fragment.

	2) $\omega \gg \lambda$. As we shall see, this case turns out to be
more actual in the case of emission from fragments. The fragments have
time to make a plenty of revolutions until the relaxation of the nuclear
surface. In this case  $\delta \to \frac\pi 4$.
The asymmetry parameter (\ref{grg7}) becomes
 \be
R(\theta)=-\frac12 a\sin 2\theta\;. \label{grg8}
\ee

\section{The results}

Simple estimations show that at average spin $I \approx 7$ the fragments
rotate many times for the time of 10$^{-19}$~s. For this reason,
the calculations were performed with the second limiting value
of $\delta = \pi/4$.
According to the full experimental spectrum \cite{peelle},
average number of quanta emitted per fission is
$N_\gamma^f\approx 8$~fission$^{-1}$. Assuming the isotropic
angular distribution of these quanta,
and given the total probability of the electric dipole quanta of (\ref{Ng})
with the angular distribution (\ref{grg2}), we can estimate the
parameter $a$ to be $a \approx 1.5\times 10^{-3}$.
By means of (\ref{grg8}) we then  get an estimation of the asymmetry
$R(\theta)\approx -0.7\times 10^{-3}$ for the angles of 35$^\circ$ and
57$^\circ$. This value is about twice as small as the experimental
values~\cite{dan} for these angles,
$R^{exp}(35^\circ$) = (1.2 $\pm$ 0.3)$\times10^{-3}$, and
$R^{exp}(57^\circ$) = (1.8 $\pm$ 0.3)$\times10^{-3}$.
In  \cite{dan} the reported $R$ values are positive. Negative sign
of our answer agrees with the intuitive picture that the both fragments have
parallel spins to each other, opposite to the angular momentum of the
relative motion of the fragments and the spin of the fissile nucleus.
Allowing for the many approximations made in calculation
\cite{dub93}, a conclusion can be made on
a satisfactory agreement with experiment. We thus summarise that
the induced polarisation mechanism of the electric dipole radiation from
fission fragments leads to a strong ROT effect with the emitted
gamma quanta. Study of this effect gives information on the dynamics of
the snapping back nuclear surface process.
Hence that is  a direct confirmation
of this phenomenon~\cite{halp}, which is of great interest, but very hardly
observed. Previous evidence of this effect was obtained in the shaking
muons emitted from the prompt fission fragments~\cite{book,D1,D2}.

\bigskip
The author is grateful to G.V.Danilyan for inducing discussions.

\end{document}